\documentclass[prl,aps,amsmath,amssymb,floatfix,showpacs,twocolumn]{revtex4}

\usepackage{graphicx,bm, amsmath, amsfonts,amssymb,amsthm}
\usepackage{subfigure}

\begin{document}

\title{Quantum Simulation of Topological Majorana Bound States and Their Universal Quantum Operations Using Charge-Qubit Arrays}
\author{Ting Mao}
\author{Z. D. Wang}
%\thanks{}
\affiliation{Department of Physics and Center of Theoretical and Computational Physics, The University of Hong Kong, Pokfulam Road, Hong Kong, China}

\date{\today}

\begin{abstract}
Majorana bound states have been a focus of condensed matter research for their potential applications in topological quantum computation. Here we utilize two charge-qubit arrays to explicitly simulate a DIII class one-dimensional superconductor model where Majorana end states can appear. Combined with one braiding operation, universal single-qubit operations on a Majorana-based qubit can be implemented by a controllable inductive coupling between two charge qubits at the ends of the arrays. We further show that in a similar way, a controlled-NOT gate for two topological qubits can be simulated in four charge-qubit arrays. Although the current scheme may not truly realize topological quantum operations, we elaborate that the operations in charge-qubit arrays are indeed robust against certain local perturbations.
\end{abstract}

   % 03.67.Lx Quantum computation architectures and implementations
                   % 85.25.Am	Superconducting device characterization, design, and modeling
                   %    Superconducting logic elements and memory devices; microelectronic circuits

\maketitle

\textit{Introduction.}---The experimental pursuit of Majorana bound states (MBSs) in one-dimensional (1D) systems has been brought into the limelight since the proposal of Kitaev's toy lattice model~\cite{Kitaev}. Based on Kitaev's original model, many experimental setups have been proposed, regarding different systems such as solid state systems including a 1D semiconducting wire on an s-wave superconductor~\cite{Lutchyn,Alicea1}, a 1D metallic wire on an unconventional superconductor~\cite{Nagaosa} and 1D quantum Ising models when a Jordan-Wigner transformation is performed, such as an array of nonlinear cavities~\cite{Bardyn} and a superconducting circuit~\cite{You}. The non-Abelian statistics of MBSs is also demonstrated by using the braiding operations realized in, for instance, a T-shaped junction~\cite{Alicea1,You} or a tunnel-coupled ancillary cavity~\cite{Bardyn}. However, braiding operations alone are not sufficient to perform universal topological quantum computation~\cite{Nayak}. Some topologically unprotected operations have to be introduced to implement other single-qubit gates~\cite{Schmidt} and two-qubit gates, e.g., a typical one---controlled-NOT (CNOT) gate~\cite{Xue}.

Superconducting-qubit circuits have been widely explored in quantum-state engineering and quantum computation due to theirs capability to control the state of a single qubit and the interqubit couplings~\cite{Makhlin}. In this paper, we employ a tunable inductively coupled charge-qubit array as the building block for topological qubits. With Jordan-Wigner transformation, we can explicitly map two charge-qubit arrays to the simplest model of the DIII class~\cite{Schnyder,Kitaev2,Zhao} 1D topological superconductor, which simply corresponds to the two copies of Kitaev's 1D toy model with different spin species~\cite{Nagaosa,Gu}. Using the Kramers doublet ground states of this DIII class model as the basis states of a topological qubit~\cite{note1}, it is demonstrated that universal single-qubit operations can be achieved by inductively coupling two charge qubits at the ends of the arrays (complemented by the braiding operations), which is conducted in a topologically protected way. Furthermore, we also show that a topological CNOT gate can be realized in four charge-qubit arrays in the same manner. Thus, we may claim that this charge-qubit array system provides an experimentally feasible scheme to simulate universal quantum gates within a subspace of topological ground states.

\textit{Charge-qubit array.}---We start with the building block of topological qubits---a charge-qubit array, as shown in Fig.~1. For each charge qubit, the superconducting island is connected to two superconducting quantum interference
device (SQUID) loops which provide additional control possibilities compared to single Josephson junctions. In this multicoupler design, the inductive coils can induce both nearest-neighbor coupling and non-nearest-neighbor coupling between qubits. As stated in Appendix A, with a deliberate design of the inductive coils, all the non-nearest-neighbor couplings can be safely neglected. The Hamiltonian of the charge-qubit array reads
\begin{eqnarray}
H &=& -1/2\sum_{i=1}^N (B_x\sigma_i^x+B_y\sigma_i^y+B_z\sigma_i^z) \notag \\
      &\quad&  -L/4\sum_{i=1}^{N-1}(I_x\sigma_i^x+I_y\sigma_i^y)(I_x\sigma_{i+1}^{x}+I_y\sigma_{i+1}^{y}), \label{ham}
      \end{eqnarray}
where
\begin{eqnarray}
B_x &=&  E_J [\cos (\chi_{l})+\cos (\chi_{r}-\phi_\text{ex})] \notag  \label{BI} \\
B_y &=&  -E_J [\sin (\chi_{l})+ \sin (\chi_{r}-\phi_\text{ex})]  \notag      \\
B_z &=&  4E_c(2n_g-1)          \\
I_x &=&  (\pi E_J/\Phi_0)[\sin(\chi_{l})-\sin(\chi_{r}-\phi_\text{ex})] \notag \\
I_y &=&  (\pi E_J/\Phi_0)[\cos(\chi_{l})-\cos(\chi_{r}-\phi_\text{ex})]  \notag
\end{eqnarray}
and $\Phi_0=hc/2e$ denotes the superconducting flux quantum.

\begin{figure}[tbp]
\subfigure{
 \begin{minipage}[b]{0.5\textwidth}

\includegraphics[width=8cm,height=2.5cm]{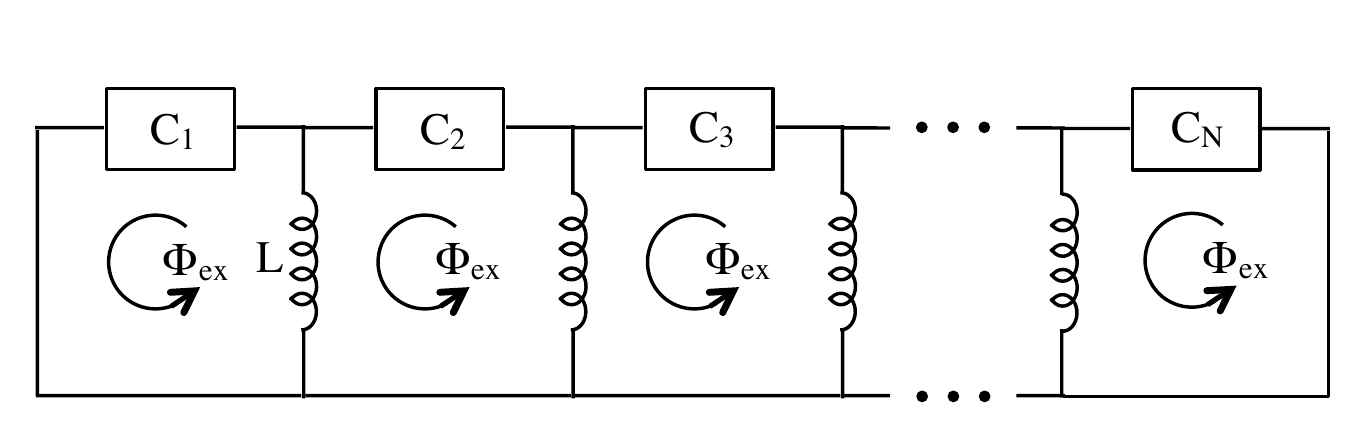}
\end{minipage}
}
\subfigure{
\begin{minipage}[b]{0.5\textwidth}
\includegraphics[width=5.5cm,height=3.2cm]{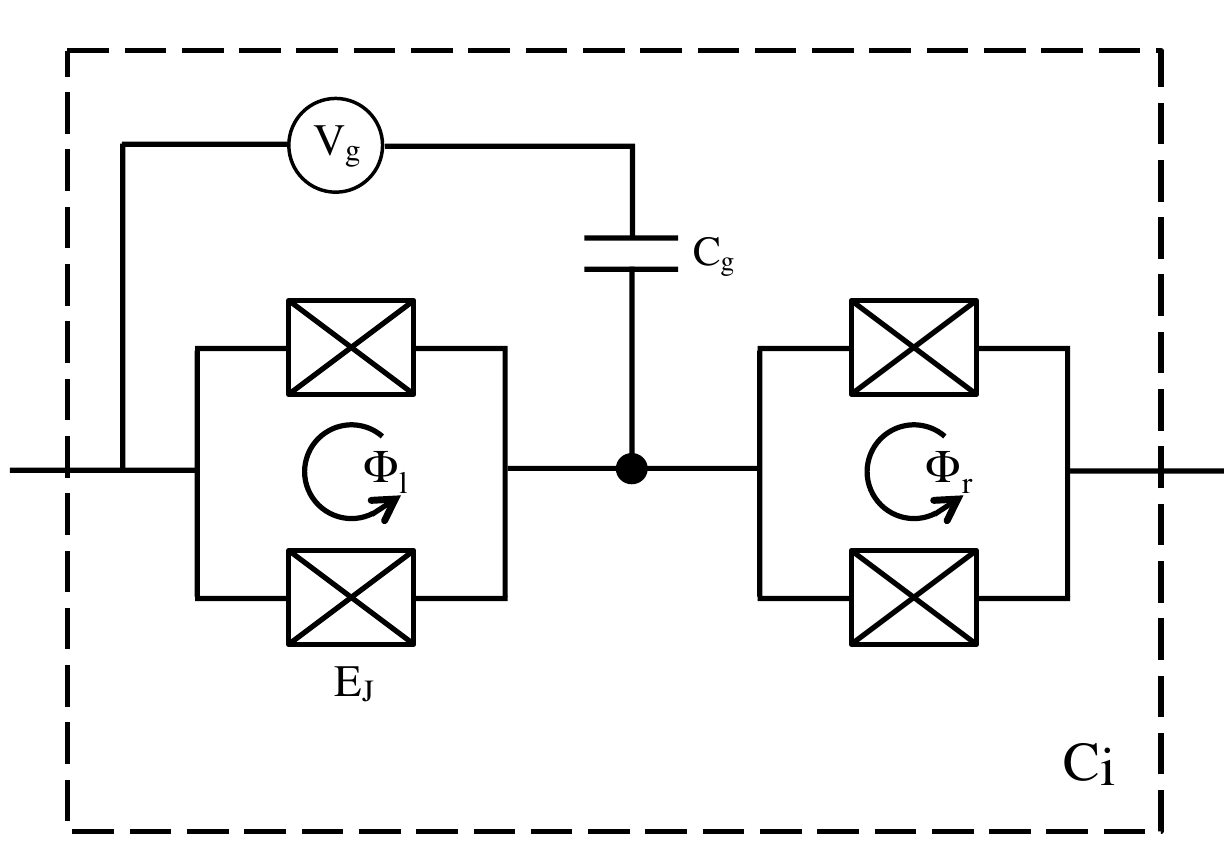}
\end{minipage}
}
\caption{ Upper panel: Charge-qubit array. Charge qubits are denoted as $C_i, i=1, 2, \dots, N$  and $\Phi_\text{ex}$ is the flux through each charge-qubit loop. Lower panel: Elements of a charge qubit. The superconducting island (denoted as a solid dot) is connected to two SQUID loops and biased by a gate voltage $V_g$ through a gate capacitance $C_g$. The two SQUIDs, biased with the fluxes $\Phi_l$ and $\Phi_r$, respectively,  consist of Josephson junctions with same Josephson energy $E_J$ and capacitance $C_J$. \label{fig1} }
\end{figure}

For clarity of discussion, we have assumed that each charge qubit has the same parameters, the inductances of the coils are all equal to $L$ and the Josephson junctions contained in the SQUID loops are identical (each with Josephson energy $E_J$ and capacitance $C_J$). The charging energy is $E_c=e^2/(2C_g+8C_J)$. The phase shifts $\chi_{l(r)}$ and $\phi_\text{ex}$ are respectively determined by the fluxes through the left (right) SQUID loop inside the charge qubit $\Phi_{l(r)}$ and the external loop $\Phi_\text{ex}$ as follows:
\begin{eqnarray}
\chi_{l(r)} &=&\arctan \big (\frac{\sin(2\pi\Phi_{l(r)}/\Phi_0)}{1+\cos(2\pi\Phi_{l(r)}/\Phi_0)} \big)
\end{eqnarray}
and $\phi_\text{ex}=2\pi\Phi_\text{ex}/\Phi_0$. The dimensionless gate charge is given by $n_g=C_gV_g/2e$ when the applied fluxes change adiabatically ~\cite{Hutter}. We consider the charge regime $E_c \gg E_J$ and use the two lowest charge states $|n=0\rangle,|n=1\rangle$ near a charge degeneracy point as the eigenstates of $\sigma^z$.

The single-qubit part of Hamiltonian ~(\ref{ham}) can be eliminated by adding some constraints on the controllable parameters $n_g$, $\chi_{l(r)}$ and $\phi_\text{ex}$. Letting $\chi_r-\phi_\text{ex}=\pi+\chi_l$, $B_x$ and $B_y$ go to zero. $B_z$ vanishes when $n_g$ is tuned to 1/2. With these constraints, we only need to consider the coupling term of Hamiltonian ~(\ref{ham}) hereafter.

\begin{figure}[tbp]
\subfigure{
 \begin{minipage}[b]{0.5\textwidth}
\includegraphics[width=6.0cm,height=3.5cm]{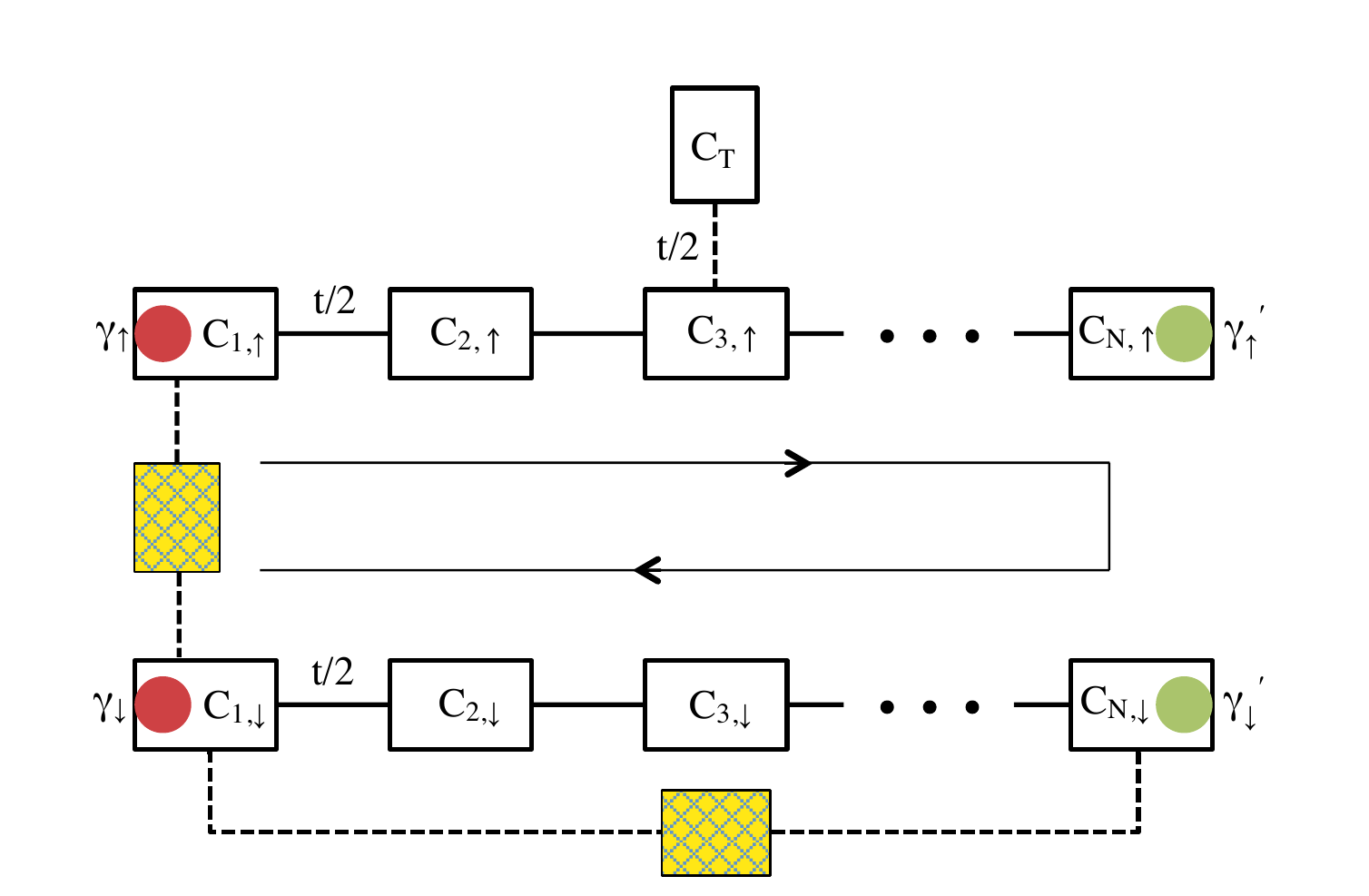}
\end{minipage}
}
\subfigure{
\begin{minipage}[b]{0.5\textwidth}
\includegraphics[width=5cm,height=2.0cm]{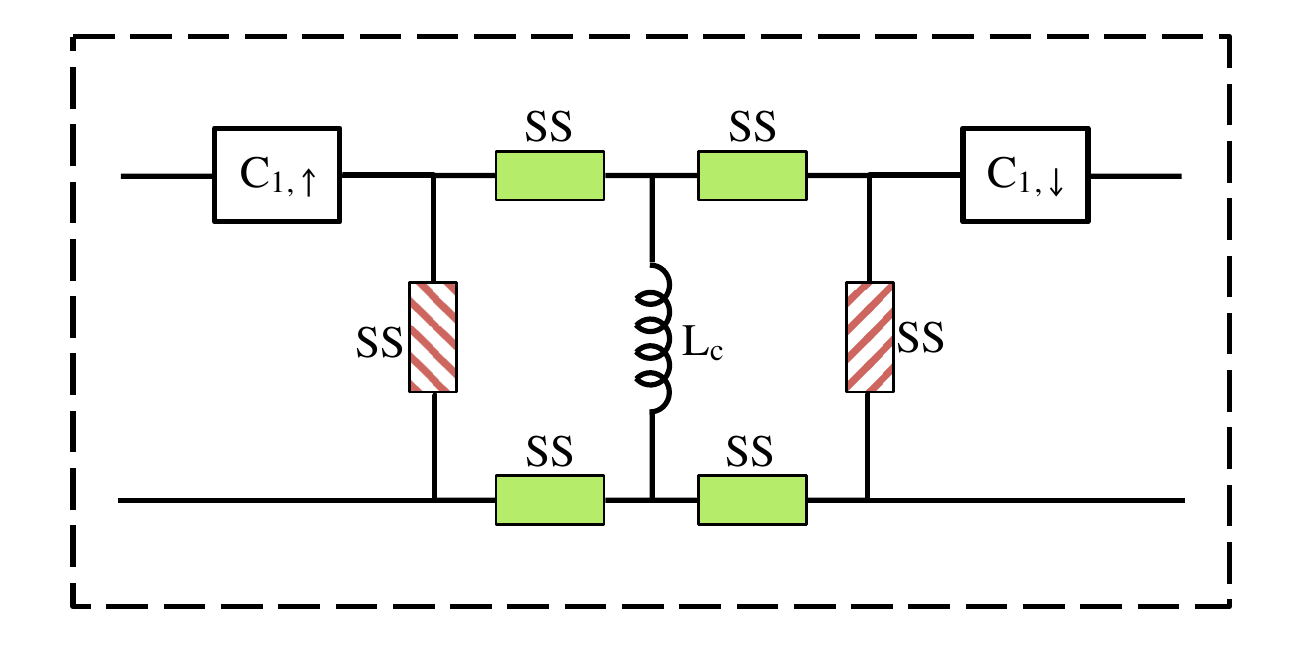}
\end{minipage}
}
\caption{(Color online) Upper panel: Two charge-qubit arrays used as a topological single qubit are schematically shown. The return path of Jordan-Wigner transformation is indicated by the line with arrows. The T-shaped junction formed by the spin-up array and an ancillary charge qubit $C_T$ can implement the braiding operation~\cite{Alicea1,You}. The controllable couplings between charge qubits at the ends of arrays are also shown. Lower panel: Controllable coupling circuit connecting $C_{1,\uparrow}$ and $C_{1,\downarrow}$. The inductive coupling is turned on (off) when the green (gray-shaded) superconducting switches (SS) are switched on (off) and the red (diagonally filled) ones off (on) at the same time. \label{fig2} }
\end{figure}
\textit{Universal single-qubit operations.}---As schematically depicted in Fig.~2, we study two such charge-qubit arrays, which are respectively denoted as spin up ($\uparrow$) and spin down ($\downarrow$) for convenience of discussions. In order to map these two arrays to a DIII class 1D topological superconductor protected by a $T^2=-1$ time reversal symmetry (TRS)~\cite{Kitaev2,Zhao}, we carefully choose a return path used for the Jordan-Wigner transformation as shown in Fig.~2. The Jordan-Wigner transformation is described as $f_{i,\uparrow}=1/2\prod_{j=1}^{i-1}(-\sigma_{j,\uparrow}^z)(\sigma_{i,\uparrow}^x-i\sigma_{i,\uparrow}^y)$ and $f_{i,\downarrow}=1/2\prod_{k=1}^{N}(-\sigma_{k,\uparrow}^z)\prod_{j=i+1}^{N}(-\sigma_{j,\downarrow}^z)(\sigma_{i,\downarrow}^x-i\sigma_{i,\downarrow}^y)$. Now using the Jordan-Wigner transformation, followed by a $U(1)$ gauge transformation of the forms $c_{i,\uparrow}=e^{i\frac{\pi}{4}}f_{i,\uparrow}$ and $c_{i, \downarrow}=e^{-i\frac{\pi}{4}}f_{i,\downarrow}$, we have the Hamiltionian of the simplest model of a DIII class topological superconductor:
\begin{eqnarray}
H &=& -t\sum_{s=\uparrow, \downarrow}\sum_{i=1}^{N-1}(c_{i,s}^{\dag}-c_{i,s})(c_{i+1,s}^{\dag}+c_{i+1,s}), \label{ham1}
\end{eqnarray}
where $c_i^{\dag}$ and $c_i$ are the creation and annihilation operators for Dirac fermions and $t=L(\pi E_J/\Phi_0)^2$ when $\chi_l$ is tuned to $\pi/4$. The TRS can be seen if we define a pseudo-TRS in the spin-up and spin-down charge-qubit arrays as follows~\cite{note2}:
\begin{eqnarray*}
T\sigma_{i,\uparrow}^xT^{-1}= -i\sigma_{i,\downarrow}^y ,& T\sigma_{i,\uparrow}^yT^{-1}=-i\sigma_{i,\downarrow}^x ,& T\sigma_{i,\uparrow}^zT^{-1}=\sigma_{i,\downarrow}^z, \notag\\
T\sigma_{i,\downarrow}^xT^{-1} = i\sigma_{i,\uparrow}^y, & T\sigma_{i,\downarrow}^yT^{-1}=i\sigma_{i,\uparrow}^x,  & T\sigma_{i,\downarrow}^zT^{-1}=\sigma_{i,\uparrow}^z.
\end{eqnarray*}
%with $TiT^{-1}=-i$.
Then the corresponding fermion operators should be transformed as
\begin{eqnarray}
Tc_{i,\uparrow}T^{-1}=P_{\uparrow}P_{\downarrow}c_{i,\downarrow}, & Tc_{i,\uparrow}^{\dag}T^{-1}=P_{\uparrow}P_{\downarrow}c_{i,\downarrow}^{\dag}, \notag \\ Tc_{i,\downarrow}T^{-1}=-P_{\uparrow}P_{\downarrow}c_{i,\uparrow}, & Tc_{i,\downarrow}^{\dag}T^{-1}=-P_{\uparrow}P_{\downarrow}c_{i,\uparrow}^{\dag} ,    \label{trs}
\end{eqnarray}
where $P_{s}\equiv \prod_{k=1}^{N}(-\sigma_{k,s}^z)
=\prod_{k=1}^{N}(1-2c_{k,s}^{\dag}c_{k,s})$, $(s=\uparrow, \downarrow)$ represents the total fermion parity of the spin-up (spin-down) chain, which will be further discussed later.

Now we can construct a topological qubit out of the degenerate ground states of the Hamiltonian.
% is commutative with.
Introducing the Majorana operators
\begin{eqnarray}
 \gamma_{i,\uparrow}=c_{i,\uparrow}+c_{i,\uparrow}^{\dag}, & \gamma_{i,\uparrow}'=i(c_{i,\uparrow}^{\dag}-c_{i,\uparrow}), \notag \\ \gamma_{i,\downarrow}=c_{i,\downarrow}+c_{i,\downarrow}^{\dag}, & \gamma_{i,\downarrow}'=i(c_{i,\downarrow}-c_{i,\downarrow}^{\dag}),
\end{eqnarray}
one can define two Dirac fermion operators $d_{\uparrow}=1/2(\gamma_{\uparrow}+i\gamma_{\uparrow}')$ and $d_{\downarrow}=1/2(\gamma_{\downarrow}-i\gamma_{\downarrow}')$ where $\gamma_s\equiv\gamma_{1,s}$ and $\gamma_s'\equiv\gamma_{N,s}'$ $(s=\uparrow, \downarrow)$ are the MBSs at the ends of the chains.

At this stage, it is notable that the analog of the MBSs in a charge-qubit system might be spatially extended due to the nonlocality of Jordan-Wigner transformation. One may be concerned that this fact could render the charge-qubit system less capable of simulating the operations on the MBSs which are strictly localized at the edges of a superconductor system. However, as elaborated in Appendix B, we illustrate that the analog of the MBSs, although it may not be topologically stable against some local disorders, could be rather stable to local perturbations in the charge-qubit system. Furthermore, it is shown that the charge-qubit system and the superconductor system, which can isomorphically be mapped to each other using Jordan-Wigner transformation, share the same ground state space. Therefore, we may conclude that the operations in the topologically protected ground state space of a superconductor system could also be simulated by the corresponding operations in the charge-qubit arrays in a relatively stable way.

Now we denote the orthogonal ground states of one Kitaev's chain as $|0\rangle_s, |1\rangle_s$ which satisfy $d_s|0\rangle_s=0$, $|1\rangle_s=d_s^{\dag}|0\rangle_s$ $(s=\uparrow, \downarrow)$. We use the convention that  $|0\rangle$ has an even and $|1\rangle$ an odd fermion parity (e.g., when the number of charge quits in one array $N$ is even). The four degenerate ground states of Eq.~(\ref{ham1}) can be denoted as: $|0\rangle_{\uparrow}|1\rangle_{\downarrow},  |1\rangle_{\uparrow}|0\rangle_{\downarrow}, |0\rangle_{\uparrow}|0\rangle_{\downarrow},|1\rangle_{\uparrow}|1\rangle_{\downarrow}$. We choose the Kramer doublet as the basis of a qubit: $\{|01\rangle, |10\rangle\}$ (where we drop the spin indices for simplicity), forming a subspace with odd fermion parity. It is noted that $P_{\uparrow}P_{\downarrow}$ always takes a value of $-1$ in such subspace if we revisit Eq.~(\ref{trs}), which guarantees the time reversal invariance of this model.

To perform universal single-qubit operations\cite{Zhu-Wang}, we start with a commonly used one, i.e., the braiding operation. For example, in a T-shaped charge-qubit junction as shown in Fig.~2, with turning on (off) the interqubit coupling $t/2$ in a certain sequence~\cite{You}, the positions of $\gamma_{\uparrow}$ and $\gamma_{\uparrow}'$ can be exchanged, generating a unitary transformation $U_z(\frac{\pi}{4})=\exp(i\frac{\pi}{4} \tau_z)$~\cite{Alicea1,Alicea2}, where $\tau_{\mu} (\mu=x,y,z)$ are the Pauli matrices in the qubit basis. We further note that the coupling term of two MBSs is commutative with the Hamiltonian~(\ref{ham1}) since the MBSs do not enter this Hamiltonian. Thus it is natural to use such kind of coupling ~\cite{Schmidt} to generate the unitary transformation of the qubit state. For instance, $U_z(\alpha)=\exp(i\alpha\tau_z)$ can be performed by turning on the coupling $i\lambda\gamma_{\uparrow}'\gamma_{\uparrow}$ for a time span $\Delta t$, where $\alpha=\lambda \Delta t$. In principle, other transformations such as $U_x$ and $U_y$ can also be achieved by observing that $i\gamma_{\downarrow}\gamma_{\uparrow}'$ and $i\gamma_{\uparrow}\gamma_{\downarrow}$ correspond to $\tau_x$ and $\tau_y$ in the qubit basis, respectively. However, only $U_y$ can be practically realized in the charge-qubit arrays. It can be deduced that an experimentally realizable coupling, i.e., the inductive coupling between two charge qubits at the ends of the arrays, is subject to not only the coupling of MBSs but also the fermion parity operators, shown as follows:
\begin{eqnarray}
(\sigma_{1,\uparrow}^x+\sigma_{1,\uparrow}^y)(\sigma_{1,\downarrow}^x+\sigma_{1,\downarrow}^y)  = & -2i P_{\uparrow}P_{\downarrow}\gamma_{\uparrow}\gamma_{\downarrow} & \sim \tau_y, \notag \\
(\sigma_{1,\downarrow}^x+\sigma_{1,\downarrow}^y)(\sigma_{N,\uparrow}^x+\sigma_{N,\uparrow}^y)  = &-2 P_{\downarrow}\gamma_{\downarrow}\gamma_{\uparrow}'  & \sim \tau_y , \notag \\
(\sigma_{1,\downarrow}^x+\sigma_{1,\downarrow}^y)(\sigma_{N,\downarrow}^x+\sigma_{N,\downarrow}^y)  =& 2i P_{\downarrow}\gamma_{\downarrow}\gamma_{\downarrow}' & \sim \tau_0    ,     \label{coupling}
\end{eqnarray}
where $P_{\uparrow,\downarrow}$ acts as $\pm \tau_z$. Now from Eq.~(\ref{coupling}), it is straightforward to see that $U_y$ can be realized simply by inductively coupling the charge qubits $C_{1,\uparrow}$ and $C_{1,\downarrow}$ \cite{note3}. $U_0$ can also be obtained in the same way which will be used in the implementation of CNOT gate. By a combination of the braiding operations $U_z(\frac{\pi}{4})$ and $U_y$, universal single-qubit operations can be achieved: $U_y(\beta)U_z(\frac{\pi}{4})U_y(\alpha)$ \cite{Schmidt}.

\begin{figure}[tbp]
\includegraphics[width=8cm,height=4.5cm]{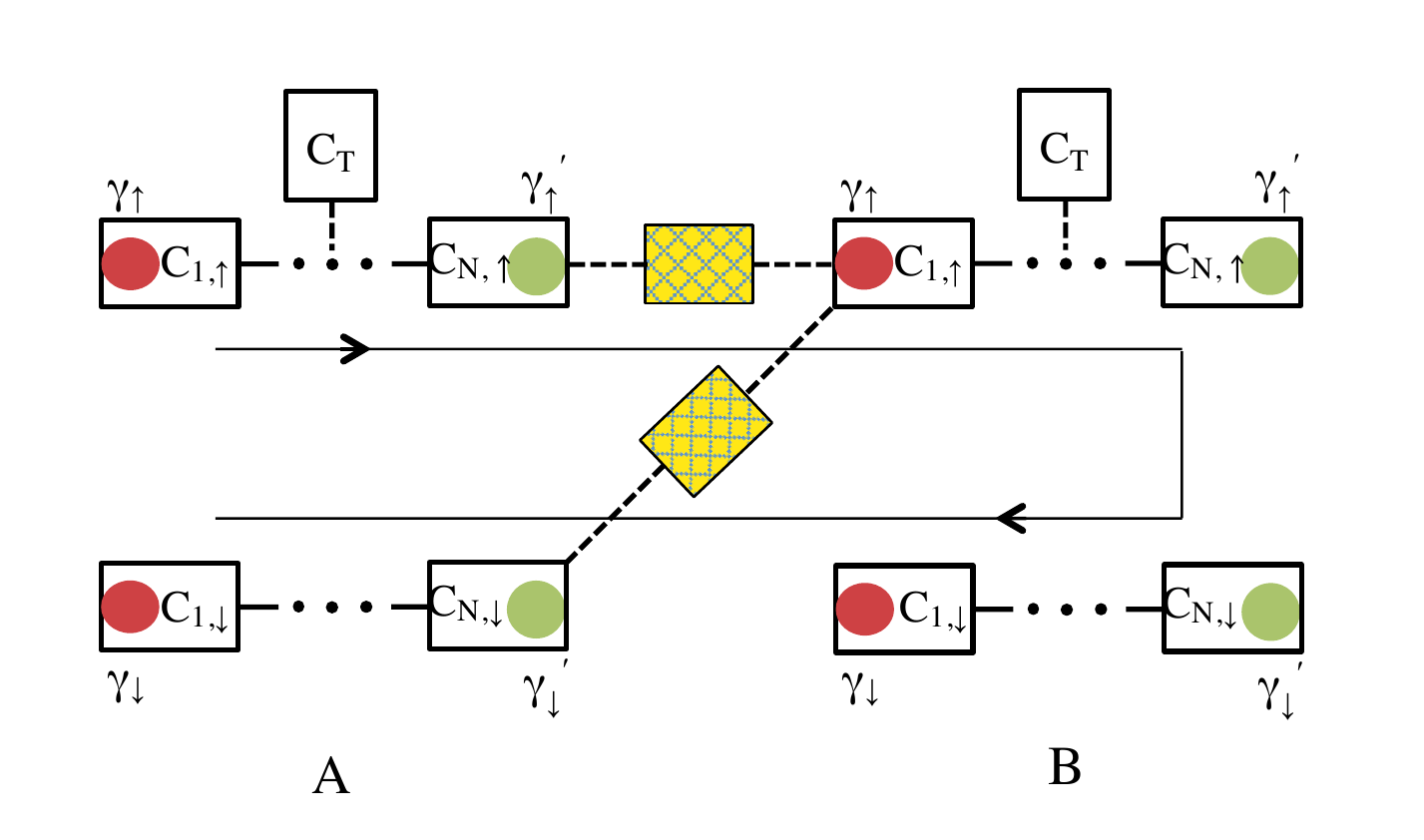}

\caption{(Color online) CNOT gate built on two pairs (A and B) of the charge-qubit arrays is schematically shown. The return path of Jordan-Wigner transformation and only the controllable couplings between the pairs A and B are shown. \label{fig3} }
\end{figure}
\textit{Controlled-NOT gate.}---The CNOT gate is the most commonly used two-bit gate and has been proved as an essential element, together with all single-qubit operations, to form a universal set which is sufficient for any quantum computation~\cite{Barenco}. Since we have already shown that one pair of charge-qubit arrays can be used as a topological qubit, it is natural to perform CNOT gate in two pairs (denoted by A and B) of the arrays, as depicted in Fig.~3. Following the same mapping as done before, we come to the basis states of two qubits $\mathcal{B}_4=\{|01\rangle_A|01\rangle_B, |01\rangle_A|10\rangle_B, |10\rangle_A|01\rangle_B, |10\rangle_A|10\rangle_B\}$. To realize the CNOT gate, it is conventional to first generate the unitary operation $U_{yy}(\alpha)=\exp(i\alpha\tau_y^A\tau_y^B)$ by turning on the Ising-type coupling $\tau_y^A\tau_y^B$ which, however, is not applicable in our system. Instead, our goal is to generate the unitary transformation $U_{0yy}(\alpha)=\exp(i\alpha\omega_0\tau_y^A\tau_y^B)$ which acts on the space spanned by the extended basis $\mathcal{B}_8=\{\mathcal{B}_4, |00\rangle_A|00\rangle_B, |00\rangle_A|11\rangle_B, |11\rangle_A|00\rangle_B, |11\rangle_A|11\rangle_B\}$ where we introduce the Pauli matrices $\omega_\mu (\mu=x,y,z)$ to represent the fermion parity space spanned by two ground states of Eq.~(\ref{ham1}), $\{|01\rangle, |00\rangle\}$ or $\{|10\rangle, |11\rangle\}$, and $\omega_0$ is the unit matrix. Following the same scheme as for single-qubit operations, we have two controllable  couplings between qubit A and qubit B (as shown in Fig.~3):
\begin{eqnarray}
(\sigma_{1,B\uparrow}^x +\sigma_{1,B\uparrow}^y)(\sigma_{N,A\downarrow}^x+\sigma_{N,A\downarrow}^y)  =& -2 P_{B}\gamma_{B\uparrow}\gamma_{A\downarrow}'  \sim \omega_y\tau_0^A\tau_x^B, \notag \\
(\sigma_{1,B\uparrow}^x + \sigma_{1,B\uparrow}^y)(\sigma_{N,A\uparrow}^x+\sigma_{N,A\uparrow}^y)  = & 2i \gamma_{B\uparrow}\gamma_{A\uparrow}'  \sim \omega_y\tau_y^A\tau_x^B ,\quad  \label{coupling2}
\end{eqnarray}
where $P_B\equiv P_{B\uparrow}P_{B\downarrow}$. Then $U_{0yy}(\alpha)$ can be obtained by a sequence of $U_{y0x}$, $U_{yyx}$ and the single-qubit operation $U_{00y}$:
\begin{eqnarray}
U_{0yy}(\alpha) &= & U_{y0x}(\frac{\pi}{2})U_{00y}(\frac{3\pi}{2})U_{yyx}(\frac{\pi}{4}) U_{y0x}(\frac{\pi}{2})\notag \\
 &\times& U_{00y}(\frac{\pi}{4})U_{y0x}(\alpha)U_{00y}(\frac{\pi}{4})U_{yyx}(\frac{\pi}{4}).
\end{eqnarray}
$U_{0yy}$ is equivalent to $U_{yy}$ in the space formed by the basis $\mathcal{B}_4$. Now it is relatively straightforward to realize the CNOT gate as follows~\cite{Li}:
\begin{eqnarray}
U_\text{CNOT} &= & U_{y0}(\frac{\pi}{4})U_{yy}(\frac{\pi}{2})U_{0z}(\frac{3\pi}{2}) U_{z0}(\frac{\pi}{4})\notag \\
 &\times& U_{yy}(\frac{\pi}{2})U_{0z}(\frac{\pi}{4})U_{yy}(\frac{\pi}{4}) U_{0z}(\frac{\pi}{4}) U_{z0}(\frac{\pi}{4}) \notag \\
   &\times& U_{x0}(\frac{-\pi}{4})U_{y0}(\frac{-\pi}{4})U_{0x}(\frac{-\pi}{4}) U_{00}(\frac{\pi}{4}).
\end{eqnarray}

\textit{Summary.}---We have shown that universal single-qubit gates and the CNOT gate performed on topological Majorana-based qubits can be simulated in the charge-qubit array system by simply controlling the inductive coupling between charge qubits, owing to the design and control flexibilities of the superconducting circuit. This scheme may also be generalized to multi-qubit cases, which is quite promising for potentially simulating the so-called topological quantum computation.

We would like to thank D. B. Zhang and Y. X. Zhao for helpful discussions. This work was supported by the GRF (HKU7058/11P\&HKU7045/13P), the CRF (HKU8/11G) of Hong Kong, and the URC fund of HKU.

\setcounter{equation}{0}
\renewcommand{\theequation}{A.\arabic{equation}}

\section{appendix a}
Here we only focus on the coupling part of the Hamiltonian and start with a simple case. First it is straightforward to write the Hamiltonian \cite{Hutter} for only two charge qubits as
\begin{eqnarray}
H_c = -\frac{L}{4} (I_x\sigma_1^x+I_y\sigma_1^y)(I_x\sigma_2^x+I_y\sigma_2^y)
\end{eqnarray}
with $I_x, I_y$ defined as in ~(\ref{BI}). In the case of three qubits, the loop in the middle can be viewed as a parallel connection of two inductive coils with a mutual inductance $M$ between them. In order to get a small value of the equivalent inductance, the two identical inductive coils  are deliberately connected in an opposite configuration. We also make an applicable assumption that the values of $M$ and $L$ are very close. Then the equivalent inductance $L_e=\frac{L^2-M^2}{2L+2M}$ would be considerably small. It is also known that as the number of connected inductive coils increases, the value of $L_e$ would be increasingly lower. Thus it is easy to know that any non-nearest-neighbor coupling between the qubits is negligible.

\setcounter{equation}{0}
\renewcommand{\theequation}{B.\arabic{equation}}
\section{appendix b}
For simplicity, we will only consider the spin-up charge-qubit array (and we drop the spin-up label hereafter)
\begin{eqnarray}
H = -2t\sum_{i=1}^{N-1} (\sigma_i^x+\sigma_i^y)(\sigma_{i+1}^x+\sigma_{i+1}^y), \label{B_ham}
\end{eqnarray}
where $t=L(\pi E_J/\Phi_0)^2$. The corresponding Hamiltonian of a topological superconductor is
\begin{eqnarray}
H &=& -t \sum_{i=1}^{N-1}(c_{i}^{\dag}-c_{i})(c_{i+1}^{\dag}+c_{i+1}). \label{B_ham1}
\end{eqnarray}
The ground state of this simple Hamiltonian can be found by constructing $|F\rangle=\prod_{i=1}^{N-1} d_i|vac\rangle$, which satisfies $d_i|F\rangle=0  (i=1,\dots,N-1)$.
The $d_i$ is defined as
\begin{eqnarray}
d_i &=& \frac{1}{2}(\gamma_i'+i\gamma_{i+1}) \\
&=& \frac{\sqrt{2}i}{4}\prod_{j=1}^{i-1}(-\sigma_{j}^z)\big[-\sigma_i^z(\sigma_{i+1}^x+
\sigma_{i+1}^y)-i(\sigma_{i}^x-\sigma_{i}^y)\big]. \notag
\end{eqnarray}
Now we introduce the representations of the $\sigma$ operators to describe the ground states, such as the two eigenstates of $\sigma^x+\sigma^y$:
\begin{eqnarray}
|\rightarrow \rangle=\frac{1}{\sqrt{2}}(1, e^{i\pi/4} )^{T},  \quad
|\leftarrow \rangle=\frac{1}{\sqrt{2}}(1, e^{i5\pi/4})^{T}.
\end{eqnarray}
We can see that a simple choice of $|F\rangle$ would be $|F\rangle=-|\leftarrow \leftarrow \dots \leftarrow \rangle$.
Then the ground state where left (right) MBS appears can be written as: $|L\rangle=\gamma|F \rangle=|\leftarrow \leftarrow \dots \leftarrow \rangle$ ($|R\rangle=\gamma'|F \rangle=|\rightarrow \rightarrow  \dots \rightarrow \rangle$).

With regard to local perturbations in the charge-qubit array, we need to consider Eq.(\ref{ham}) and some constraints of the parameters, e.g., $\chi_r-\phi_\text{ex}=\pi+\chi_l$ and $\chi_l=\pi/4$. After some deduction, we know that there are five different types of possible local perturbations in the array: $-4E_C(\delta n_g^i)\sigma_i^z$, $\frac{\sqrt{2}}{4}E_J(\delta \chi_l^i)\sigma_i^x$, $\frac{\sqrt{2}}{4}E_J(\delta \chi_l^i)\sigma_i^y$, $t(\delta \chi_l^i)(\sigma_i^x-\sigma_i^y)(\sigma_{i+1}^x+\sigma_{i+1}^y)$ and $t(\delta \chi_l^{i+1})(\sigma_i^x+\sigma_i^y)(\sigma_{i+1}^x-\sigma_{i+1}^y)$. It can be seen that each of these perturbations tends to drive the ground state ($|L\rangle$ or $|R\rangle$) to an excited state which has an energy gap $\Delta$ above the ground state. For example, since there are $\sigma^z|\rightarrow \rangle=|\leftarrow \rangle$ and $\sigma^z|\leftarrow \rangle=|\rightarrow \rangle$, we have $\Delta=8t$ for $1<i<N$, and $\Delta=4t$ for $i=1$ and $N$ when $-4E_C(\delta n_g^i)\sigma_i^z$ acts on the ground states. Such state transition is not permitted for a small perturbation like $-4E_C(\delta n_g^i)\sigma_i^z$. The same argument can be straightforwardly applied to the other four types of perturbations. Thus, we can see that under the protection of the energy gap, the ground states $|L\rangle$ and $|R\rangle$ where MBSs appear are stable to local perturbations.

We should note that $|L\rangle$ and $|R\rangle$ are also the two %orthogonal
degenerate ground states of Eq.(\ref{B_ham}). %which, indicates that Ham.(\ref{B_ham}) and (\ref{B_ham1}) share the same ground state space.


\begin{thebibliography}{99}
\bibitem{note2}
In principle, we want to have the spin-up (-down) hard-core boson operators transformed in a similar way as the fermion operators do in Eq.~(\ref{trs}) under TRS, e.g., $b^{\dag}_{i,\uparrow}=\sigma^x_{i,\uparrow}+i\sigma^y_{i,\uparrow} \rightarrow -b^{\dag}_{i,\downarrow}=-(\sigma^x_{i,\downarrow}+i\sigma^y_{i,\downarrow})$ and $b_{i,\uparrow}=\sigma^x_{i,\uparrow}-i\sigma^y_{i,\uparrow} \rightarrow b_{i,\downarrow}=\sigma^x_{i,\downarrow}-i\sigma^y_{i,\downarrow}$. This implies that $\sigma^x$ and $\sigma^y$ exchange their roles under TRS, e.g.,$\sigma^x_{i,\uparrow} \rightarrow \sigma^y_{i,\downarrow}$ and $\sigma^y_{i,\uparrow} \rightarrow -\sigma^x_{i,\downarrow}$, which, can be implemented by adding an extra $\pi/2$ flux into the the $\chi_l$ flux of the spin-down array.
\bibitem{note3}

A possible scheme to realize a controllable inductive coupling between two charge qubits (e.g.,$C_{1,\uparrow}$ and $C_{1,\downarrow}$) is to use a superconducting circuit containing one inductive coil (its inductance $L_c$ should be chosen smaller than $L$ in order to avoid the fusion of two MBSs $\gamma_{\uparrow}$ and $\gamma_{\downarrow}$) and several superconducting switches as shown in Fig.~2. The ability of turning on (off) the coupling is provided by the superconducting switches which allow switching between insulating and superconducting states in a very short switching time under the control of external field. We can see that this scheme is fully conducted in a topologically protected way. It should be also noted that since $L_c$ used is smaller than $L$, the corresponding mutual inductance $M_c$ should be adjusted to ensure that we only need to consider the coupling between the two charge qubits at the ends of arrays.
\end{thebibliography}
\end{document}